\documentclass{revtex4}

\usepackage{enumerate}
\usepackage{enumitem}
\usepackage{graphicx}
\usepackage{dcolumn}
\usepackage{bm}
\usepackage{graphicx}
\usepackage{amssymb}
\usepackage{epstopdf}
\usepackage{latexsym}

\usepackage[english]{babel}
\newcommand{\LTO}{$\mbox{LaTiO}_3${$\mbox{ } $}}
\newcommand{\STO}{$\mbox{SrTiO}_3${$\mbox{ }  $}}
\newcommand{\LAO}{$\mbox{LaAlO}_3${$\mbox{ }  $}}
\newcommand{\TiO}{$\mbox{TiO}_2${$\mbox{ }  $}}
\begin{document}

\title{Irreversibility and time relaxation in electrostatic doping of oxide interfaces.}

\author{J. Biscaras$^1$, S. Hurand$^1$, C. Feuillet-Palma$^1$, A. Rastogi$^2$, R. C. Budhani$^{2,3}$, N. Reyren$^4$, E. Lesne$^4$, D. LeBoeuf$^5$, C. Proust$^5$, J. Lesueur$^1$, N. Bergeal$^1$}

\affiliation{$^1$LPEM- UMR8213/CNRS - ESPCI ParisTech - UPMC, 10 rue Vauquelin - 75005 Paris, France}
\affiliation{$^2$Condensed Matter - Low Dimensional Systems Laboratory, Department of Physics, Indian Institute of Technology Kanpur, Kanpur 208016, India}
\affiliation{$^3$National Physical Laboratory, New Delhi - 110012, India }
\affiliation{$^4$Unit\'e Mixte de Physique CNRS-Thales, 1 Av. A. Fresnel, 91767 Palaiseau, France }
\affiliation{$^5$Laboratoire National des Champs Magn\'etiques Intenses - Toulouse, 143 avenue de Rangueil, 31400 Toulouse, France }

\date{\today}

\maketitle

 Two-dimensional electron gas (2DEG) confined in quantum wells at insulating oxide interfaces \cite{Ohtomo:2004p442} have attracted much attention as their electronic properties display a rich physics with various electronics orders such as superconductivity \cite{Reyren:2007p214, Biscaras:2010p7764, Perna} and magnetism\cite{bert:2011p767,li:2011p762,brinkman:2007p493}. A particularly exciting features of these hetero-structures lies in the possibility to control their electronic properties by electrostatic gating \cite{Caviglia:2008p116,biscaras2}, opening up new opportunities for the development of oxide based electronics\cite{hwang, bibes,Takagi:2010p9802,Mannhart:2010p6675}. However, unexplained gating hysteresis and time relaxation of the 2DEG resistivity have been reported in some bias range, raising the question of the precise role of the gate voltage. Here we show that in \LTO/\STO and \LAO/\STO heterostructures, above a filling threshold, electrons irreversibly escape out of the well. This mechanism, which is directly responsible for the hysteresis and time relaxation, can be entirely described by a simple analytical model derived in this letter.  Our results highlight the crucial role of the gate voltage both on the shape and the filling of the quantum well. They also demonstrate that it is possible to achieve a low-carrier density regime in a semiconductor limit,  whereas the high-carrier density regime is intrinsically limited. \\

In oxides hetero-structures such as \LTO/\STO or \LAO/\STO, the bending of the \STO conduction band forms a quantum well at the interface where a 2DEG takes place.  Figure 1a shows the numerical simulation of the conduction band profile and sub-bands filling obtained by solving self-consistently Schr\"{o}dinger-Poisson equations \cite{biscaras2,Meevasana:2011p9019}. As opposed to semiconducting hetero-structures where the electrons fill only the bottom of the potential well, here, the Fermi energy is  intrinsically  located near the top of the well. This results from  a peculiar situation where the bending of the conduction band is mostly determined by the electron spatial distribution itself. At low temperature, the extension of the gas is of order of 5 nm corresponding to approximatively 12 units cell of \STO (figure 1) \cite{basletic,Copie:2009p5635}.  The 2DEG accommodates mainly  two types of carriers : the majority low-mobility carriers (LMC) that fill the sub-bands deep in the well  and the minority high-mobility carriers (HMC) located in the upper part of the well \cite{biscaras2}.  The presence of the two types of carriers is only observed  below a temperature of approximately 70~K, which corresponds to the range where the dielectric constant $\epsilon(V,T)$ of \STO strongly increases when the temperature is lowered and becomes electric field dependent\cite{NEVILLE:1972p3397,biscaras2}.  The LMC are confined close to the interface where  $\epsilon(V,T)$ is strongly reduced by the electric field, whereas the HMC extend more in \STO where $\epsilon(V,T)$ tends to recover its buk value \cite{biscaras2}.\\

We have studied \LTO and \LAO epitaxial layers grown on \TiO -terminated  \STO single crystals by Pulsed Laser Deposition as described in the Supplementary Material section.  A metallic back-gate is deposited at the rear of each 0.5 mm thick \STO substrate. The main text of this article focuses mostly on  the results obtained on the \LTO/\STO heterostructure whereas the data concerning the  \LAO/\STO heterostructure can be found in the Supplementary Material section.  We start by cooling the \LTO/\STO heterostructure to 4.2K without applying  gate voltage (i. e. un-gated). The evolution of the carrier density $n_\mathrm{Hall}= \frac{B}{eR_\mathrm{Hall}}$ with temperature, extracted from Hall effect measured at low magnetic field (7T) is shown in figure 2a. The drastic fall of  $n_\mathrm{Hall}$  below $\sim$70~K corresponds to the separation of the 2DEG into two types of carriers, as revealed by the observation of a non-linear Hall effect at higher magnetic field (figure 2a inset). The total carrier density $n_\mathrm{Tot}=n_\mathrm{LMC}+n_\mathrm{HMC}$  shown in figure 2a has been extracted from a two-carrier analysis of the Hall effect that considers carriers with different densities and mobilities ($n_\mathrm{LMC}$, $\mu_\mathrm{LMC}$) and  ($n_\mathrm{HMC}$, $\mu_\mathrm{HMC}$) \cite{biscaras2,Ohtsuka:2010p9619}. In the high temperature regime $n_\mathrm{Tot}=n_\mathrm{Hall}$, whereas $n_\mathrm{Tot}\neq n_\mathrm{Hall}$ in the low temperature regime (T$<$70K). The overall decrease of the carrier density from 300K to 4K is due to the trapping of electrons in the defects of  \STO.

After cooling the sample, the transport properties of the heterostructure are investigated when a positive voltage is applied for the first time on the gate, a situation referred as the first positive polarization. After a small decrease, the resistance of the 2DEG reaches a saturation value that no longer evolves with gate voltage (figure 2b). In a simple Drude picture, such behaviour would indicate that the carrier density remains constant, although, according to electrostatic laws, one expect electrons to be added to the 2DEG. This first polarization is irreversible, since, when the gate voltage is decreased from its maximum value $V_\mathrm{G}^{\mathrm{max}}$=200~V, the resistance curve deviates from the previous one. Similar behaviour has been mentioned in the literature \cite{Bell:2009p6086,Reyren:2007p214}. On the same figure, we also plot the evolution of the carrier densities $n_{\mathrm{LMC}}$, $n_{\mathrm{HMC}}$ and $n_{\mathrm{Tot}}$=$n_{\mathrm{LMC}}+n_{\mathrm{HMC}}$ with gate voltage, extracted from a two-carrier analysis of  the high field Hall effect.  The total number of carriers $n_\mathrm{Tot}$ is found to be rather constant during the first polarization, which is consistent with the saturation of the resistance. The slight decreases of the LMC density is compensated by a slight increase of  the HMC one.  As already observed in the resistance curve, when the gate voltage is decreased, the total number of carrier does not follow the first polarisation curve.\\

 Figure 3a shows the filling of the quantum well for different values of $V_\mathrm{G}$ that illustrates the previous experimental observations.  The effects of the gate voltage on  the interface is twofold : (i) it adds ($\Delta V_\mathrm{G}>0$) or removes ($\Delta V_\mathrm{G}<0$) electrons to the 2DEG, (ii) it controls the shape of the upper part of the  well by tilting the conduction band profile in the substrate. For positive gate voltage, this two effects take place simultaneously to produce HMC as sub-bands are filled near the top of the well, where the band profile becomes shallower. As shown in figure 1a, HMC are also intrinsically present in un-gated hetero-structures during the first cool down, confirming that the Fermi level is located in the upper region of the well. Therefore, when the gate voltage is increased towards positive values, the added electrons rapidly fill the sub-bands all the way up to the top of the well to finally escape in the conduction band of the \STO substrate. Hence, the total carrier density and the resistance of the sample saturate. This behaviour is only possible because the tilting of the conduction band of the  \STO  has a negative slope away from the interface. The evolution of the carriers mobilities  $\mu_{\mathrm{LMC}}$ and $\mu_{\mathrm{HMC}}$ during the first positive polarisation is consistent with the previous observations (inset figure 2b).  Indeed, the mobility of LMC confined deep in the well is found to be rather constant. On the other hand,  the mobility of HMC increases rapidly for positive gate voltage before saturating, indicating that the electron added fills the last sub-bands accessible before escaping from the well.\\

The irreversibility of the first positive polarisation has been investigated in more detail. After cooling down an un-gated sample  to 4.2~K, its sheet resistance is measured as a function of gate voltage for different polarization procedures (figure 3b). As opposed to the previous measurements,  the first polarisation is made towards negative voltage down to $V_{\mathrm{G}}^\mathrm{min}$=-200~V. The resistivity increases as expected when electrons are removed from the 2DEG, and no saturation is observed. As the voltage is brought back to  $V_{\mathrm{G}}$=0~V, the resistance curve appears to be fully reversible. However, when the gate voltage is further increased to the positive value of $V_{\mathrm{G}}^\mathrm{max1}$=50~V before being decreased back to $V_{\mathrm{G}}$=-200~V  the curve deviates from the previous one.  On the other hand, this new curve  is fully reversible as long as the gate voltage is not increased above $V_{\mathrm{G}}^\mathrm{max1}$. The same pattern has been repeated for increasing maximum values of $V_{\mathrm{G}}$ ($V_{\mathrm{G}}^\mathrm{max2}$=100~V, and $V_{\mathrm{G}}^\mathrm{max3}$= 200~V). The same behaviour is also observed in the case of the \LAO/\STO sample (supplementary figure 3). These results confirms the scenario proposed previously. The well can be emptied ($V_{\mathrm{G}}\searrow$) or filled ($V_{\mathrm{G}}\nearrow$) reversibly as long as the gate voltage do not increase above a critical value corresponding to the maximum value $V_{\mathrm{G}}^\mathrm{max}$ previously applied to the metallic gate. Beyond this value, the electrons escape irreversibly into the \STO substrate and are lost for the 2DEG. In the literature, a first positive polarisation to the highest voltage value is often used prior to the other measurements to suppress the hysteresis effects \cite{Caviglia:2008p116,biscaras2}. \\
 
The previous experiment raised the question of the escape mechanism of the electrons in the substrate. We performed time resolved resistivity measurements to study the response of the 2DEG to a gate voltage step $\Delta V_{\mathrm{G}}\pm$10~V. Typical results of these measurements are shown in the insets of figure 3b.  When the voltage steps occur in a reversible situation, the resistance shows a clear jump $\Delta R$ before reaching a stable value, $\Delta R$ being negative for $\Delta V_{\mathrm{G}}>0$ and positive for $\Delta V_{\mathrm{G}}<0$ (left inset figure 3b). On the contrary, measurements performed in an irreversible situation with a final value of  $V_{\mathrm{G}}$ exceeding the previous $V_{\mathrm{G}}^\mathrm{max}$, display a negative jump of the resistance, before it recovers slowly its original value (right inset figure 3b).  At first approximation, this relaxation is logarithmic in time, of the form  $\alpha + \beta\log(t)$ (figure 3b). In this experiment,  the speed of the initial jump in resistance is determined by the characteristic $R_GC$ time at which electrons are added to the 2DEG (Supplementary Material). \\

To analyse the relaxation in the irreversible regime  we have developed a model that describes the dynamic of electrons escaping out the well. Details of the calculation are given in the Supplementary Material section. We model  the quantum well of figure 1a as a 2D well with an infinite barrier on the \LTO side and a barrier of finite height $E_B$ on the \STO side. The total number of electrons in the well right after a gate voltage step $\Delta V_\mathrm{G}$ is given by $n_{0^+}=n_{0^-}+C_a(V_G)\Delta V_{\mathrm{G}}/e$ where $n_{0^-}$  is the number of electrons before the voltage step and $C_a$ is the capacitance per unit of area of the \STO substrate. Assuming that the electrons at the Fermi level $E_{\mathrm{F}}$  can thermally jump across the barrier of energy $E_B$ with a first order kinetics (figure 5a), the Drude resistivity in the relaxation regime is found to be (see Supplementary Material section)
\begin{equation}
R(t) = R_{0^+}\Big[1+\frac{N(E_F)k_BT}{n_{0^+}}\ln\big(1+ \frac{t}{t_E}\big)\Big]\\ 
\label{relax}
\end{equation}
where  $N(E_F)$ is the density of states at the Fermi level and $R_{0^+}=\frac{1}{e\mu n_{0^+}}$ ($\mu$ is the electron mobility). The escape time $t_E$ that appears in equation (\ref{relax}) is given by 
\begin{equation}
t_E=\frac{k_BTN(E_F)}{\nu n_0^+}\exp\Big[\frac{E_B-E_F^{0+}}{k_BT}\Big]
\end{equation}
 where $E_F^{0+}$ is the Fermi energy after the voltage step and $\nu$ is a characteristic frequency. 
 A systematic study of the dependence of the relaxation on the polarization parameters has been carried out to validate this model. In particular, we measured the relaxation of the sheet resistance after a step $\Delta V_G$=+10V performed at different values of  $V_\mathrm{G}$ (figure 4a), and  the relaxation of the sheet resistance for different steps values $\Delta V_\mathrm{G}$=5, 10, 20, 40V (figure 4c). In both cases,  the experimental data are in very good agreement with the theoretical expression (\ref{relax}), confirming that the model catches the essential physics of the phenomena. The same agreement between experiment and theory is also observed for the \LAO/\STO sample (supplementary figure 3). To understand the dependence of $t_E$ with $V_{\mathrm{G}}$ and $\Delta V_G$, we can simply express the logarithm of the escape time as
 \begin{equation}
   \ln t_E=\gamma -\kappa C_a(V_{\mathrm{G}})\Delta V_\mathrm{G}
   \label{lnte}
 \end{equation}  
 where $\gamma$ and $\kappa$ are constants whose expressions can be found in Supplementary Material. 
   For a constant step $\Delta V_\mathrm{G}$ in the irreversible region, the variation of $\ln t_E$ is  consistent with the evolution of the capacitance $C_a$  with gate voltage that can be measured experimentally (figure 4b) \cite{biscaras2}. The variation of $\ln t_E$ with $\Delta V_G$ is found to be linear for small $\Delta V_\mathrm{G}$.  For larger $\Delta V_\mathrm{G}$, $\Delta R$ tends to saturate since when the Fermi level has reached the top of the conduction band, electrons  are almost directly injected in the \STO conduction band and spill into the substrate. \\

  As already shown in the inset of figure 3b in the limit $t\gg t_E$, the expression (\ref{relax}) reduces to $R(t)=\alpha + \beta \log{t}$ with $\alpha=R_{0^+}(1-\frac{N(E_F)k_BT}{n_{0^+}}\ln t_E)$ and $\beta=\frac{N (E_F)k_BTR_{0^+}}{2.3n_{0^+}}$.  We have verified experimentally that the parameter $\beta$ increases linearly with temperature (figure 5c). This is a direct consequence of a thermally activated mechanism. We emphasise here that the electrons escaping in the conduction band of the \STO substrate are attracted by the positive voltage of the back-gate electrode, but  can not reach it. Indeed, the substrate being a thick wide gap insulator, the electron density is too small to create a metallic state. Therefore,  electrons diffuse in the \STO substrate until they get trapped in the defects of the crystal. Hence,  these electrons do not contribute to electronic transport, but do play a role to maintain the polarisation between the two sides of the substrate. Figure 5b shows that the trapped electrons can be released in the 2DEG if the temperature of the sample is increased above two characteristic values $T_1\approx$ 70K and $T_2\approx$ 170 K in the case of the \LTO/\STO sample.  \\

In summary, we have shown that in \LAO/\STO and \LTO/\STO interfaces, the Fermi level is instrinsically close to the top of the quantum well created by the bending of the conduction band at the interface. When the carrier density is increased by an electrostatic back-gate voltage beyond a critical value, most of the added electrons escape in the \STO substrate conduction band at a rate well explained by a thermally activated leaking out of the well. This phenomenon which appears to be common to all \STO based interfaces, is directly responsible for the saturation of the 2DEG  properties with gate voltage such as the mobility and the carrier density but also the superconducting transition temperature observed at lower temperature (supplementary figure 4). It  also certainly affects the strength of Rashba spin-orbit coupling that was shown to increase strongly in the region of positive gate voltage \cite{caviglia2}. Hence, these results put a limitation on the performances achievable with back gated heterostructures,  in particular in terms of high-mobility and  Rashba spin-orbit, two important properties that originally contributed to the emergence of \STO based interface, as they open future perspectives in the field of oxides electronics. To overcome this problem we suggest to realise double gated structures. A back gate could be used to engineer the shape of the quantum well that determines the mobility of carriers through the bending of the \STO conduction band. Independently, electrons could be added into the well thanks to a top gate.\\

The Authors gratefully thank  M. Grilli, S. Caprara and A. Millis for stimulating discussions. This work has been supported by the R\'egion Ile-de-France in the framework of CNano IdF and Sesame program. Part of this work has been  supported by Euromagnet II. The work at IIT Kanpur has been funded by the Department of Information Technology. RCB acknowledge the J C Bose Fellowship of the Department of Science and Technology, Government of India\\

\newpage
\begin{figure}[h!]
\includegraphics[width=12cm]{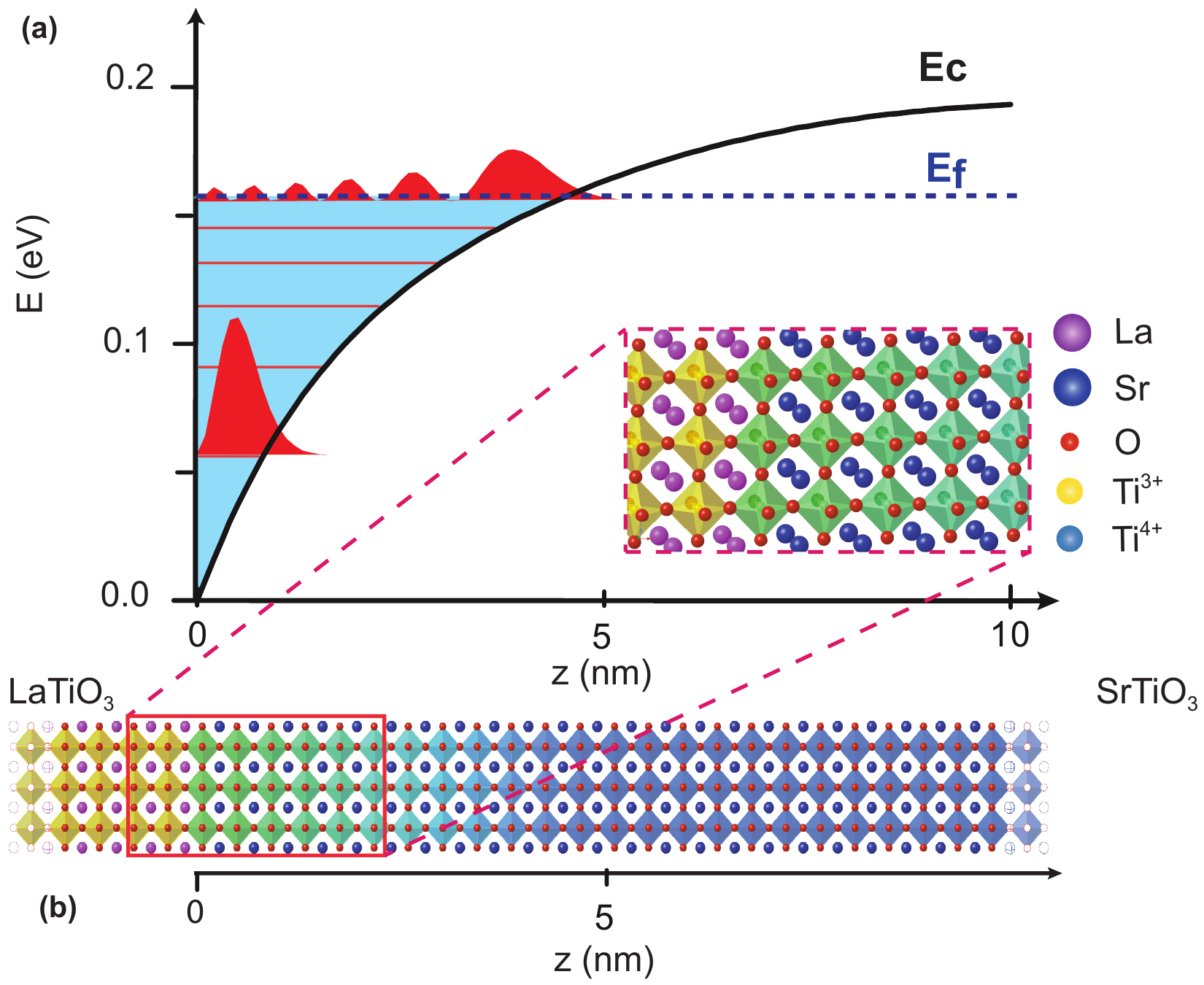}
\caption{2DEG at the \LTO/\STO interface. (a) Theoretical calculation of the quantum well profile at the \LTO/\STO interface for an un-gated sample of carrier density 7.23 10$^{13}$/cm$^2$. The figure displays the conduction band profile $E_{\mathrm{C}}$ (black), the Fermi energy $E_{\mathrm{F}}$ (green dashed), the sub-bands energies (red) as a function of depth $z$ from the interface. The square modulus of the envelope function of the first and last filled sub-bands are indicated in arbitrary units (red areas). (b) Schematic description of the crystallographic structure of the interface.The sub-band of highest energy extend up to 12 u.c in \STO whereas the sub-band of lowest energy is confined within 3 u.c. .}
\end{figure}
\newpage

\begin{figure}[h!]
\includegraphics[width=10cm]{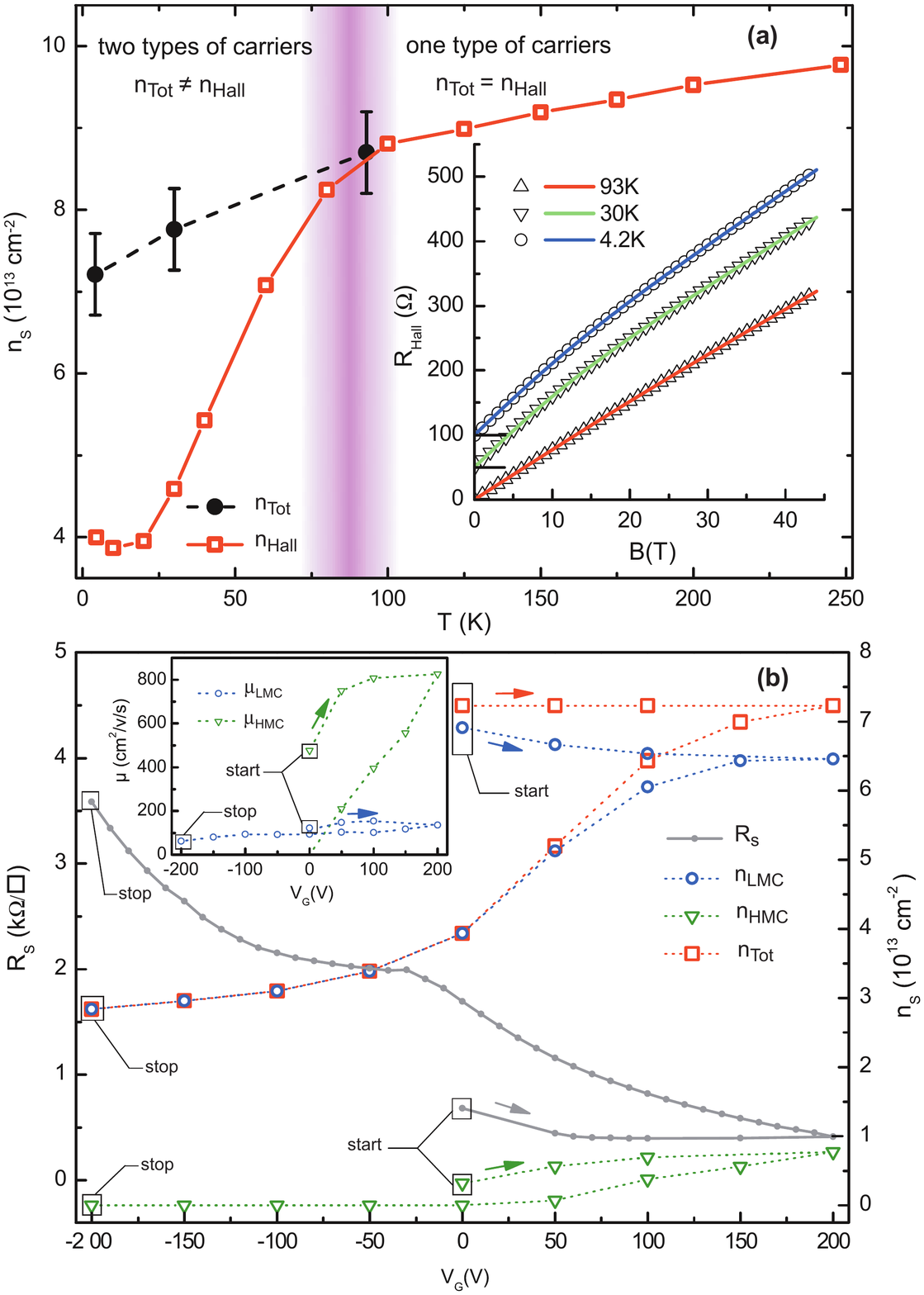}
\caption{Hall effect and first polarisation. (a) Low magnetic  field ($ <$ 7 T) Hall carrier density $n_{\mathrm{Hall}}$ as a function of temperature  and total carrier density $n_{\mathrm{Total}}$ extracted from the two-carrier model analysis of high magnetic field measurements (45 T) at three relevant temperatures. Inset : Hall resistance measured at  high field for three different temperatures.  An offset, indicated by a black horizontal segment has been added to separate the curves. (b) Resistance of the 2DEG (left scale) and carrier densities $n_\mathrm{LMC}$, $n_\mathrm{HMC}$ and $n_\mathrm{Tot}$ (right scale) as a function of gate voltage measured for the first forward sweep (first positive polarisation) and backward sweep. The starting points, the stoping points and the directions of the sweeps are indicated on the graph. Inset) Mobilities $\mu_\mathrm{LMC}$ and $\mu_\mathrm{HMC}$ of the two types of carriers measured for the same gate sweeps as in (b).}
\end{figure}

\newpage
\begin{figure}[h!]
\includegraphics[width=10cm]{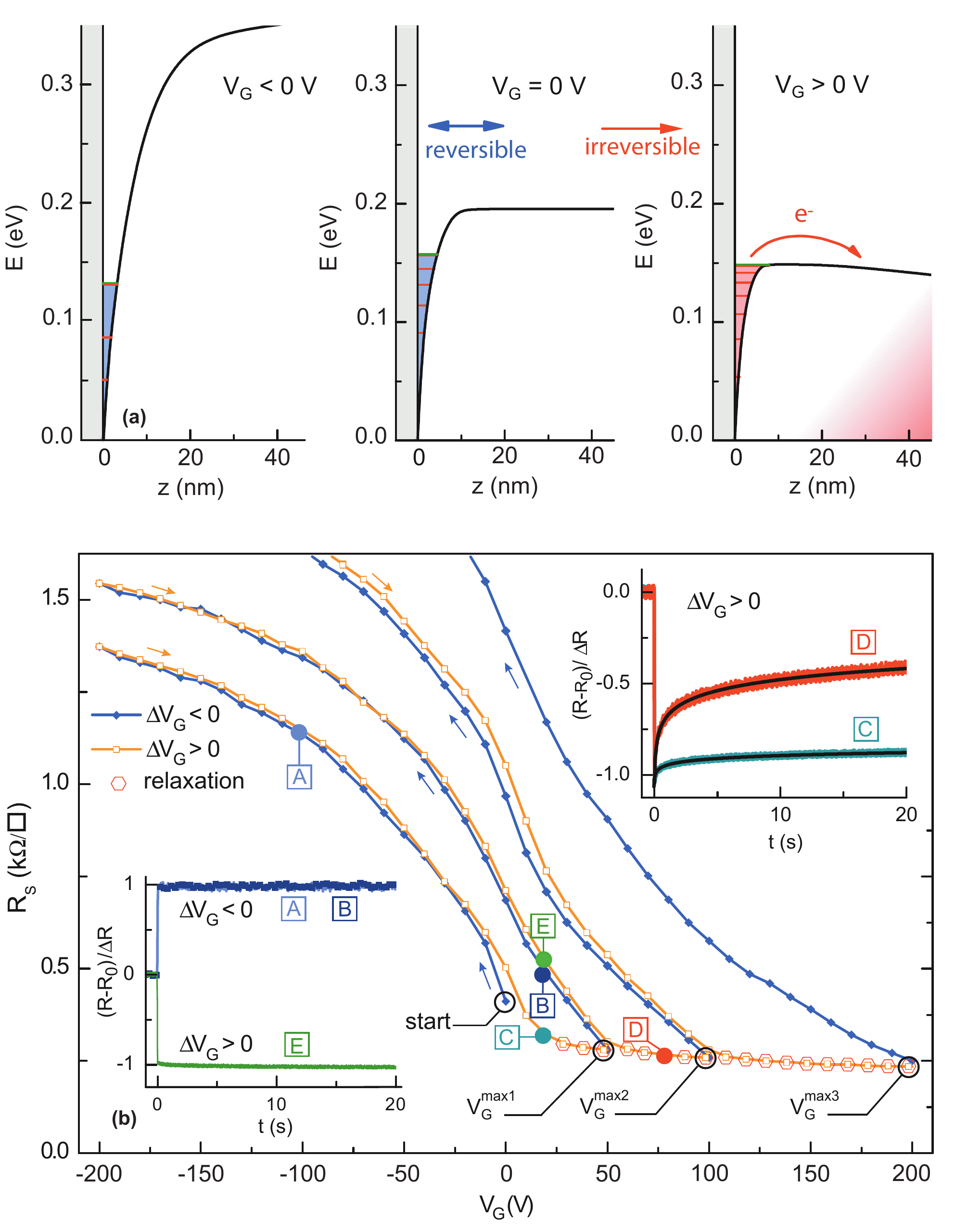}
\caption{Reversible and irreversible regimes. (a) Filling of the quantum well at the \LTO/\STO interface for different $V_{\mathrm{G}}$. Starting from $V_{\mathrm{G}}=0$, the well can be emptied and filled reversibly for negative gate voltages $V_{\mathrm{G}}<0$. By contrast, during the first positive polarisation, electrons irreversibly escape from the well. (b) Resistance of the sample measured for several forward sweeps ($\Delta V_{\mathrm{G}}>0$) and backward sweeps ($\Delta V_{\mathrm{G}}<0$) of $V_{\mathrm{G}}$ starting from the un-gated situation (``start'' point). $V_{\mathrm{G}}$ is first swept to  -200~V  and then increased to a maximum value $V_{\mathrm{G}}^\mathrm{max1}$ before being swept back to -200~V. The operation is repeated with several maximum values of the gate voltage  $V_{\mathrm{G}}^\mathrm{max2}$ and $V_{\mathrm{G}}^\mathrm{max3}$. Red markers indicate the region where relaxation is observed. Insets) Normalised resistance measured as a function of time after a  $\Delta V_{\mathrm{G}}=\pm$ 10~V step in different points of the $R(V_{\mathrm{G}})$ curves labelled ``A '',  ``B '' , ``C '' , ``D '' and ``E '' in panel b).  Curves  ``C ''  and ``D ''   are fitted by a logarithmic relaxation of the form $\alpha + \beta\log(t)$ (full black lines). }
\end{figure}

\newpage

\begin{figure}[h!]
\includegraphics[width=10cm]{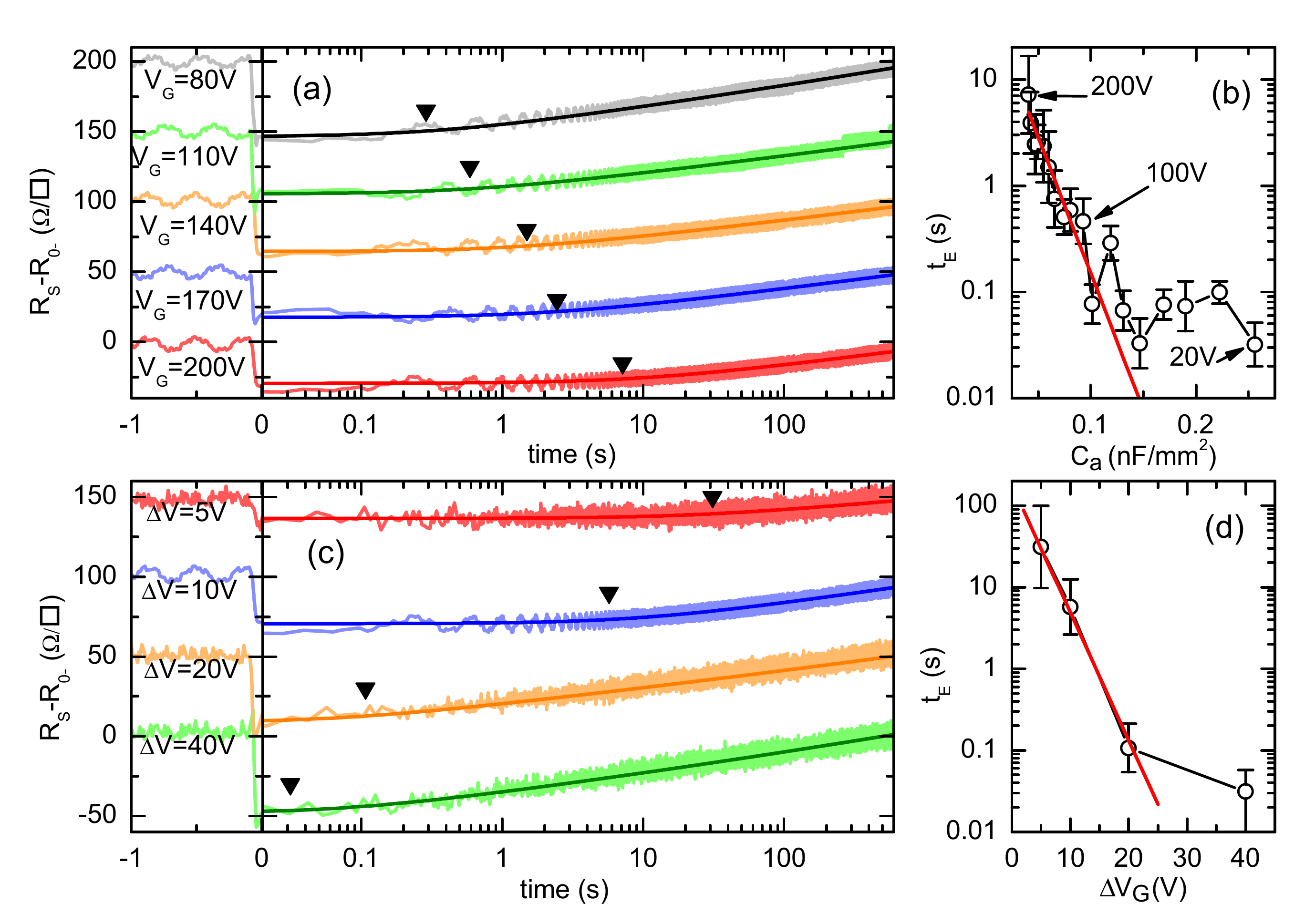}
\caption{Relaxation of the 2DEG resistance. (a) Resistance as a function of time  measured at 4.2 K for different gate voltage after a $\Delta V_G$=+10~V step, fitted by expression (\ref{relax}). Arrows indicate $t_E$ extracted from the fits. (b) Logarithm of $t_E$, plotted as a function of $C_a$ and  fitted with expression ($\ref{lnte}$) (red line). (c) Resistance as a function of time measured at 4.2 K for different steps $\Delta V_{\mathrm{G}}$, fitted by expression (\ref{relax}). Arrows indicate $t_E$ extracted from the fits. (d) Logarithm of $t_E$  plotted as a function of $\Delta V_{\mathrm{G}}$ and fitted with expression ($\ref{lnte}$). }
\end{figure} 

\newpage

\begin{figure}[h!]
\includegraphics[width=10cm]{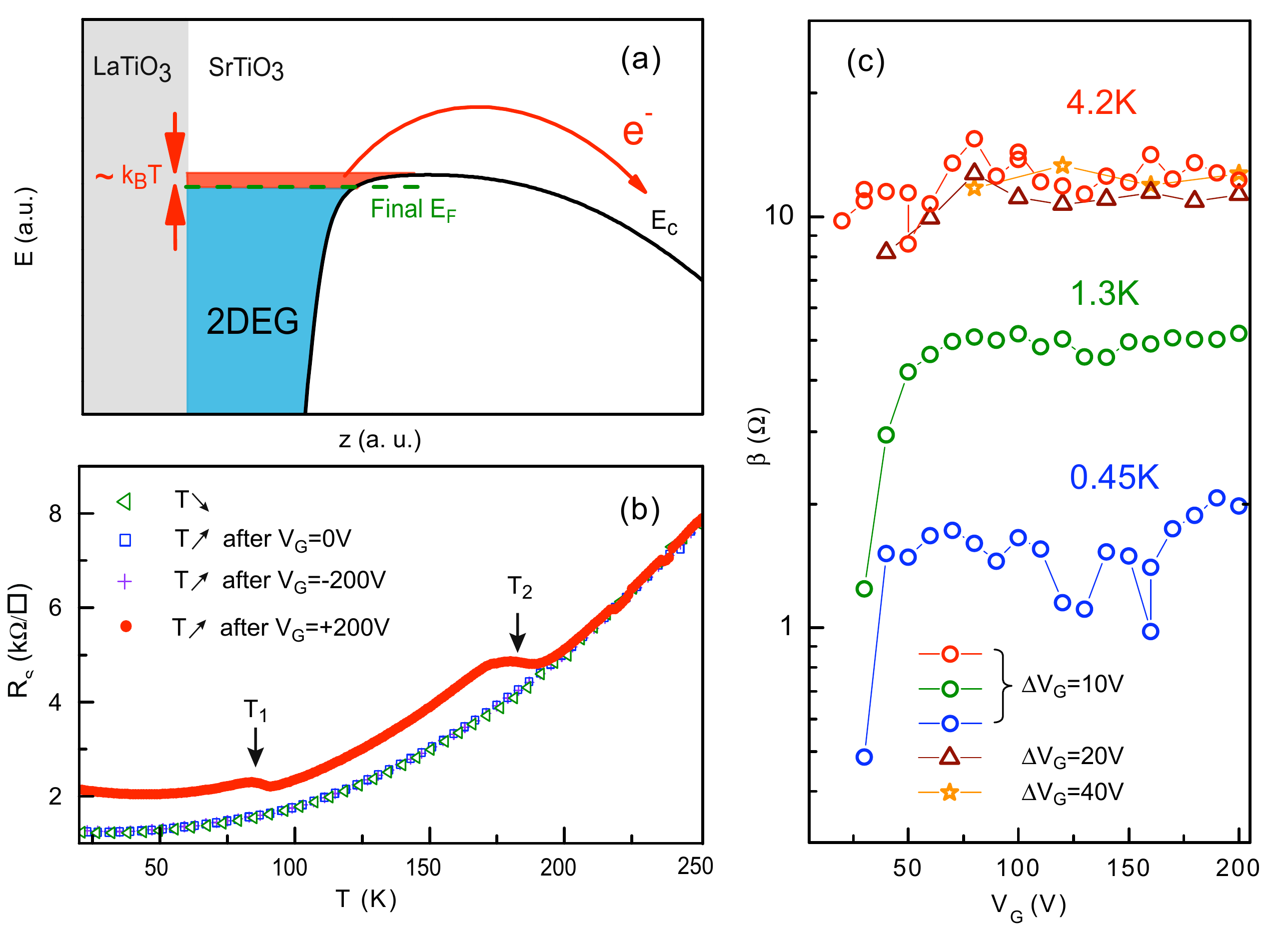}
\caption{ Thermal activated hoping and de-trapping. (a) Schematic representation of electrons escaping out of the well by thermally activated hoping. (b) Resistance as a function of temperature measured for different sweep procedures : cool down (triangle) and warm up  (square) with $V_\mathrm{G}$=0 (square), warm up with $V_\mathrm{G}$=0 after a sweep to $V_\mathrm{G}$=-200V (cross) and warm up with $V_\mathrm{G}$=0 after a sweep to $V_\mathrm{G}$=+200V (circle). (c) Parameter $\beta$ of the logarithmic fit of the relaxation curves, as a function of gate voltage for three different temperature and different $\Delta V_G$.}
\end{figure}    

\newpage

\large{\textbf{Supplementary Material}}\\
\normalsize

1. \textbf{Growth of the heterostructures}\\

$\mbox{LaTiO}_3$/$\mbox{SrTiO}_3$ hetero-structures were grown at ITT Kanpur (India)  using
excimer laser based PLD on commercially available (Crystak gmbh Germany) single crystal substrates of $\mbox{SrTiO}_3$ 
 (100)  oriented. The substrates were given a buffered HF treatment to expose $\mbox{TiO}_2$ terminated
surface. Before deposition, they were heated in oxygen pressure of 200 mTorr in the temperature range of 850 to 950 C$^\circ$ for one hour
to realize surface reconstruction.  The source of $\mbox{LaTiO}_3$  is a stoichiometric
sintered target of 22 mm in diameter which was ablated in oxygen partial
pressure of $1\times10^{-4}$ Torr with energy fluence of 1\,J/cm$^2$ per pulse
at a repetition rate of 3 Hz to acheive a gowth rate of 0.12 \,\AA/s.  Under these conditions, the $\mbox{LaTiO}_3$  phase is grown on $\mbox{SrTiO}_3$ substrates, as shown by X-Rays diffraction patterns \cite{Biscaras:2010p7764}. In this study, we used 15 u.c. thick $\mbox{LaTiO}_3$ layers on 0.5 mm thick  $\mbox{SrTiO}_3$ substrates.\\

$\mbox{LaAlO}_3$/$\mbox{SrTiO}_3$heterostructures were fabricated at UMR CNRS/Thales (Paris, France). A thin \LAO film is deposited by PLD (Surface PLD system) on a $\mbox{TiO}_2$-terminated $(0\,0\,1)$-oriented \STO substrate (Crystec and SurfaceNet). A buffered HF treatment followed by annealing, as described in Ref. \cite{Koster}, was used to obtain the $\mbox{TiO}_2$ termination required to obtain the conducting electronic system at the interface. The KrF excimer (248\,nm) laser ablates the single-crystalline \LAO target at 1\,Hz, with a fluence between 0.6 and 1.2\,J/cm$^2$ in an O$_2$ pressure of $2\times 10^{-4}$\,mbar. The substrate is typically kept at 730C$^\circ$ during the growth, which is monitored in real-time by RHEED. As the growth occurs layer-by-layer, it allows us to control the thickness at the unit cell level. After the growth of the film, the sample is cooled down to 500C$\circ$ in $10^{-1}$\,mbar of O$_2$, where the oxygen pressure is increased up to 400\,mbar. In order to reduce as much as possible the presence of oxygen vacancies (in both the substrate and the film), the sample stays in these conditions for 30 minutes before being cooled down to room temperature \cite{Cancellieri2010}. The substrate-target distance is about 57\,mm, leading to a growth rate of about 0.2 \,\AA/s in the above conditions. In this study, we used 5 u.c. thick $\mbox{LAlO}_3$ layers on 0.5 mm thick  $\mbox{SrTiO}_3$ substrates.\\

1. \textbf{Characteristic charging time of the sample}\\

\begin{figure}[h]
\includegraphics[width=10cm]{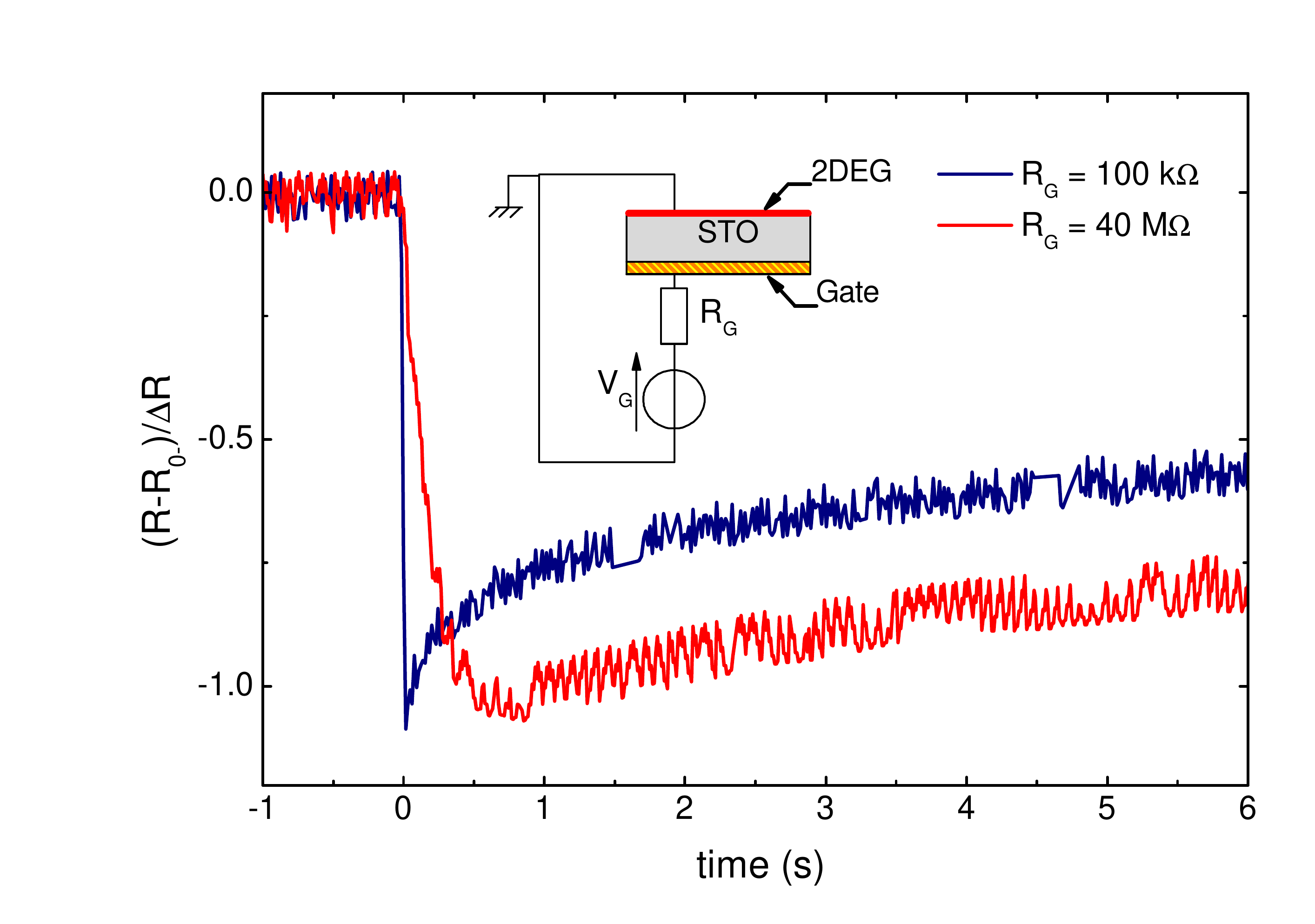}
\caption{Relaxation of the normalised resistance of the 2DEG after a gate voltage step $\Delta V_\mathrm{G} = $ 10~V  for two different external resistances $R_G$. The characteristic time $\tau = R_G C$ is un-resolved for $R_G$=100~k$\Omega$. The electrical scheme in inset shows the RC circuit formed by the heterostructure and the external resistance.}
\label{fs1}
\end{figure}

\indent The 2DEG carrier density at the interface  can be electrostatically tuned by a metallic gate deposited at the back of the \STO substrate. The heterostructure forms a capacitor, whose plates consist of the metallic gate on one side and the 2DEG on the other side, and whose dielectric material is the \STO substrate. 
As shown by the scheme in Figure~\ref{fs1}, the heterostructure of capacitance $C$ is placed in serial with an external resistor of resistance $R_G$  defining a simple   RC circuit. Therefore, the change in the charge density of the 2DEG after a voltage step is not instantaneous, but reach it's final value with a time dependence  $\mathrm{exp}(-t/\tau)$ where $\tau= R_GC$. In a simple Drude model, the resistance of the 2DEG after a voltage step is expected to evolve with the same time dependence. Figure \ref{fs1} show the resistance of the 2DEG as a function of time, upon applying a voltage step $\Delta V_\mathrm{G} = $ 10~V, for two different values of the external resistance. The blue curve corresponds to the short $\tau= R_GC$ time ($R_G=$100~k$\Omega$) used for all the experiments reported in the main text. It shows a sharp decrease of resistivity, followed by the logarithmic relaxation discussed in the main text. On the other hand, the red curve corresponds to a longer $\tau= R_GC$ time ($R_G=$40~M$\Omega$) and shows that the initial decrease of the resistance is much  slower. However, the logarithmic relaxation remains basically unchanged. We thus see that in our experiment the $\tau= R_GC$ time at which the electrons are added to the 2DEG  do not interfere with the thermal leaking of electrons into the \STO substrate. \\

2. \textbf{Thermal escape of electrons from the quantum well}\\

\begin{figure}[h]
\includegraphics[width=12cm]{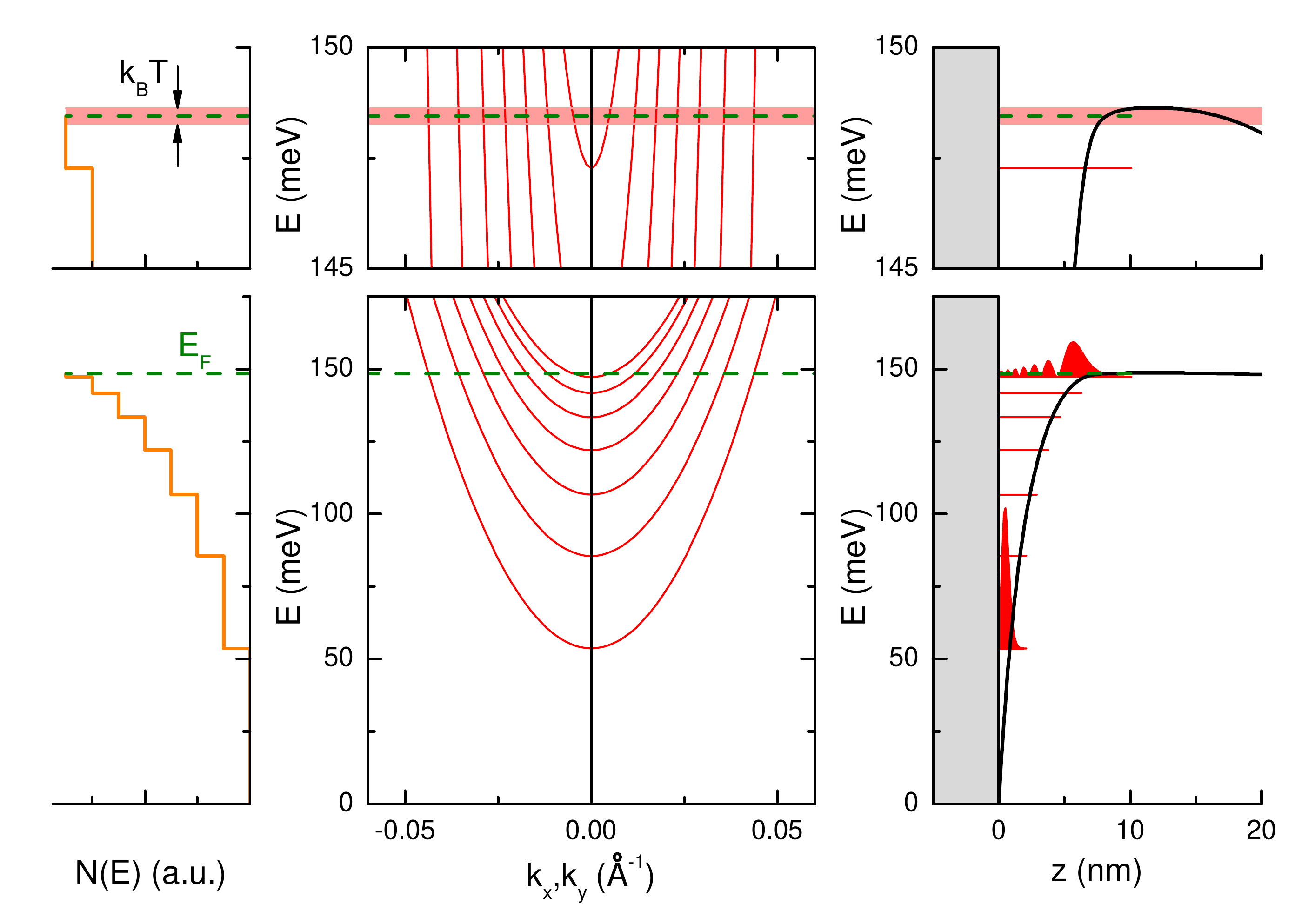}
\caption{Energy diagram of the quantum well at the \LTO /\STO interface in the irreversible regime. Left : Density of states in unit of $m / \pi \hbar^2 $. Middle : dispersion of the electronic sub-bands in the ($k_x$ , $k_y$) plane. Right : bending of the conduction band (black) in the $z$-direction. Energy levels and the first and last envelope function are plotted in red. Down : full energy scale. Top : blow-up near the maximum of the band. The pink horizontal band represent the thermal energy at 4.2~K, centered around the Fermi level (green dashed line). }
\label{fs2}
\end{figure}

\indent We have built a model to account for the relaxation of resistivity in the irreversible regime based on the thermal escape of the electrons over the top of the conduction band profile at positive gate voltage. The band bending and sub-bands energies of the \LTO /\STO interface in the irreversible regime are shown in Figure~\ref{fs2} as an illustration \cite{biscaras2}. However, this model is very general and does not depend on the details of the well such as its shape.\\

\indent We consider a 2D potential well with a finite barrier height $E_B$ and a finite number of parabolic sub-bands of energy $E_L^i$ ($i$={1,...,$n_L$}). We assume that electrons at the Fermi level $E_{\mathrm{F}}$ can thermally jump over the barrier with a first order kinetics :
\begin{equation}
\frac{\mathrm{d}n}{\mathrm{d}t} = - k n\\
\end{equation}
where $n$ is the density of electrons in the well and $k$ is the kinetic factor. For thermally activated hoping the latter follows an Arrh\'enius law : $k = \nu e^{-\frac{\Delta}{k_{\mathrm{B}} T}}$ where the activation energy is  $\Delta = E_B-E_{\mathrm{F}}$, and $\nu$ is a characteristic frequency factor. Hence, considering that the density of state of a 2D sub-band is constant and that the Fermi level always stay above the highest sub-band level (the well is almost totally filled up), the equation on $E_{\mathrm{F}}$ is :
\begin{equation}
\frac{\mathrm{d}E_{\mathrm{F}}}{\mathrm{d}t} = - \nu e^{-\frac{E_B-E_{\mathrm{F}}}{k_{\mathrm{B}} T}} (E_{\mathrm{F}}-\frac{1}{n_L}\sum_{i=1}^{n_L} E_L^i)\\
\end{equation}
which is equivalent to a problem with only one level $E_L=\sum_{i=1}^{n_L} E_L^i/n_L$ but with a density of state multiplied by $n_L$. Considering low temperatures ($k_B T << E_{\mathrm{F}}-E_L$), the variations of $E_{\mathrm{F}}$ can be neglected in $(E_{\mathrm{F}}-E_L)$ at first order, which gives the solution :
\begin{equation}
E_{\mathrm{F}}(t)=E_{\mathrm{F}}^{0+}-k_B T \mathrm{ln} \left(1+ \frac{t}{t_E} \right)
\end{equation}
where $E_{\mathrm{F}}^{0+}$ is the Fermi level at $t$=0$+$ (in our case : just after the voltage step) and $t_E$ is the characteristic escape time :
\begin{equation}
t_E = \frac{k_B T}{\nu(E_{\mathrm{F}}^{0+}-E_L)}e^{\frac{E_B-E_{\mathrm{F}}^{0+}}{k_{\mathrm{B}} T} }\\
 = \frac{k_B T N(E_{\mathrm{F}})}{\nu n_{0+}}e^{\frac{E_B-E_{\mathrm{F}}^{0+}}{k_{\mathrm{B}} T} }
\label{eqtE}
\end{equation}

where $N(E_{\mathrm{F}})= {n_L m}/{\pi \hbar^2}$ is the total density of state at the Fermi level.\\
It follows that the Fermi level is constant for $t<t_E$ and  then decreases logarithmically at longer time. Hence, the conductivity calculated from a simple Drude model with a constant mobility $\mu$ (small variation of $n$) is given by

\begin{equation}
\sigma(t) = {e \mu \left[ {n_{0+} - N(E_{\mathrm{F}}) k_B T \mathrm{ln}\left(1+ \frac{t}{t_E}\right) }\right]}\\
\end{equation}

The pre-factor of the logarithm is of the order of $10^{11}$~cm$^{-2}$, which is sufficiently small compared to the density of electrons to allow for a first order approximation in the resistivity :

\begin{equation}
R(t) = R_{0+} \left[ 1 + \frac{N(E_{\mathrm{F}}) k_B T}{n_{0+}} \mathrm{ln} \left(1+\frac{t}{t_E}\right) \right]
\end{equation}
where $R_{0+}=(e\mu n_{0+})^{-1}$ is the Drude resistivity just after the step. As shown in the main text, this time dependence of the resistance fits the experimental data with a very good accuracy, and the logarithm pre-factor shows experimentally the thermal activation of the process.\\
\\
\indent To analyse the experimental behaviour of the escape time $t_E$ with gate voltage, we now consider a positive gate voltage step $\Delta V_\mathrm{G}$ at $t=0$. Starting from a density of electrons $n_{0-}$, the density after the step is :
\begin{equation}
n_{0+} = n_{0-} + \frac{1}{e} \int_{V_\mathrm{G}^{0-}}^{V_\mathrm{G}^{0+}} C_a(V_\mathrm{G}) \mathrm{d}V
\end{equation}
where $C_a(V_\mathrm{G})$ is the capacitance per unit area of the sample. One assumption made here is that there is enough  ``space'' in the well to receive all the added electrons at initial time (which is true for small $\Delta V_\mathrm{G}$). For small $\Delta V_\mathrm{G}$,  the variations of $C_a$ with $V_\mathrm{G}$ can be neglected, such that the Fermi level rises up according to :
\begin{equation}
E_{\mathrm{F}}^{0+} = E_{\mathrm{F}}^{0-} + \frac{C_a(V_\mathrm{G}) \Delta V_\mathrm{G}}{e N(E_{\mathrm{F}})} 
\end{equation}
Hence, when replaced in equation (\ref{eqtE}), the variation of $t_E$ with $C_a$ and $\Delta V_\mathrm{G}$ is given by:
\begin{equation}
\mathrm{ln} (t_E) = \gamma - \kappa C_a(V_\mathrm{G}) \Delta V_\mathrm{G}
\label{eqlntE}
\end{equation}
where $\gamma = \mathrm{ln} \left( \frac{N(E_{\mathrm{F}}) k_B T}{\nu n_{0+}} \right) + \frac{E_B-E_{\mathrm{F}}^{0-}}{k_B T}$ and $\kappa= \frac{1}{e N(E_{\mathrm{F}}) k_B T}$.
As shown in figure 4 of the main text, the variations of $\mathrm{ln} (t_E)$ follow a linear dependence with both $C_a$ for $V_\mathrm{G} \gtrsim $ 100~V (deep irreversible regime), and $\Delta V_\mathrm{G}$  (for values lower than  40~V) at $V_\mathrm{G} = $ +200~V for instance.
\\

\begin{figure}[h]
\includegraphics[width=10cm]{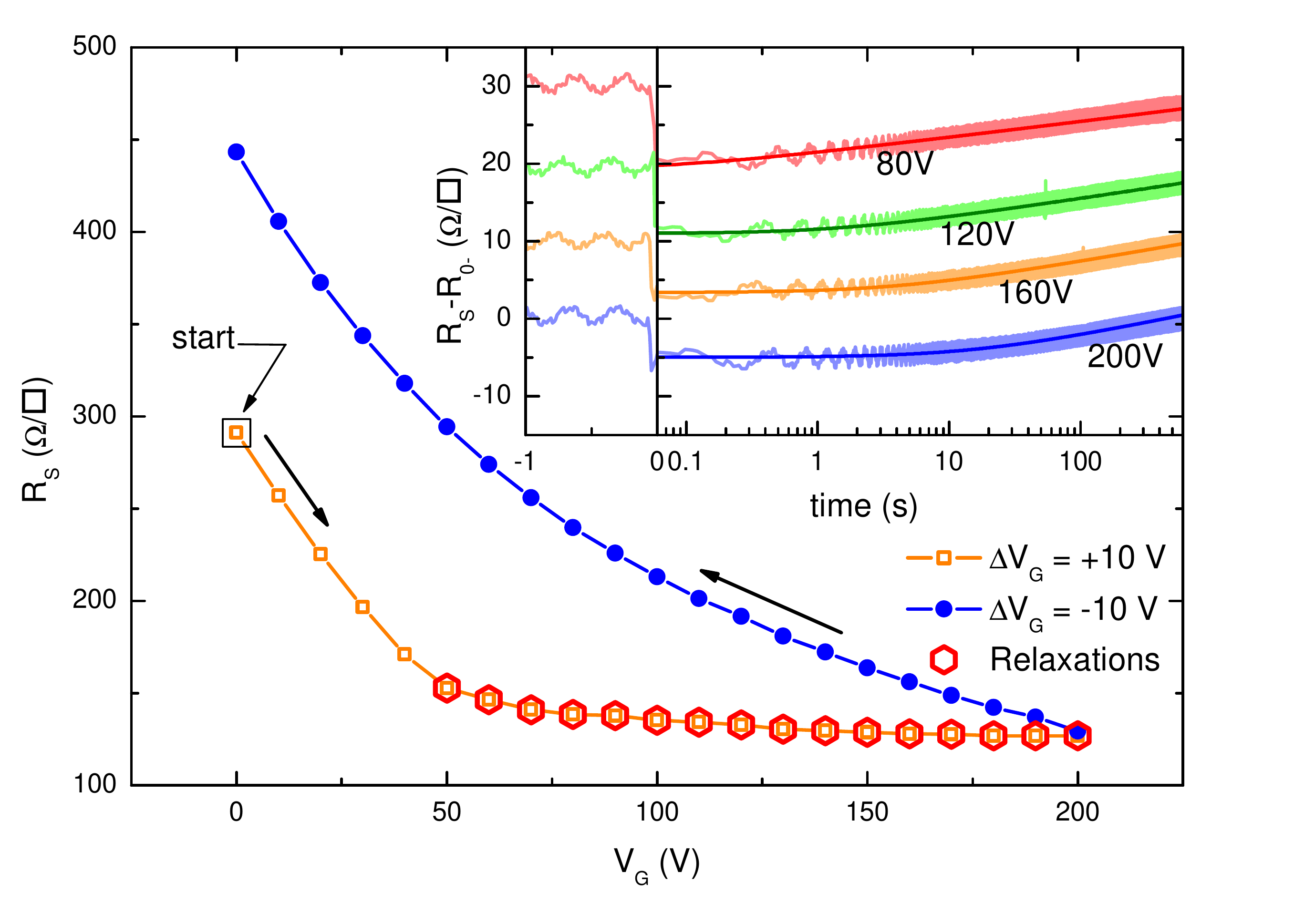}
\caption{Measurement of the resistivity of the \LAO/\STO sample during a first positive polarisation at 4.2~K. Inset : time resolved resistivity measurements (light coloured curve) during the first polarisation for selected gate voltages showing thermal escape of the electrons. An offset has been added to each curve for sake of clarity. The data are fitted by the thermal escape model (dark coloured curve).}
\label{fs3}
\end{figure}

3. \textbf{Relaxation results on \LAO/\STO interface}\\

Similar time resolved resistance measurement were carried on a 5 unit cell \LAO grown on \TiO-terminated  \STO substrate. As for the \LTO/\STO heterostructures, measurements at low temperature with different polarisation procedures  show that negative first polarisations are reversible, while the positive first polarisations are irreversible. Supplementary Figure 3 shows the resistivity saturation upon the first positive polarisation and the associated thermal escape of the electrons, which is well fitted by our model.\\

\begin{figure}[h]
\includegraphics[width=10cm]{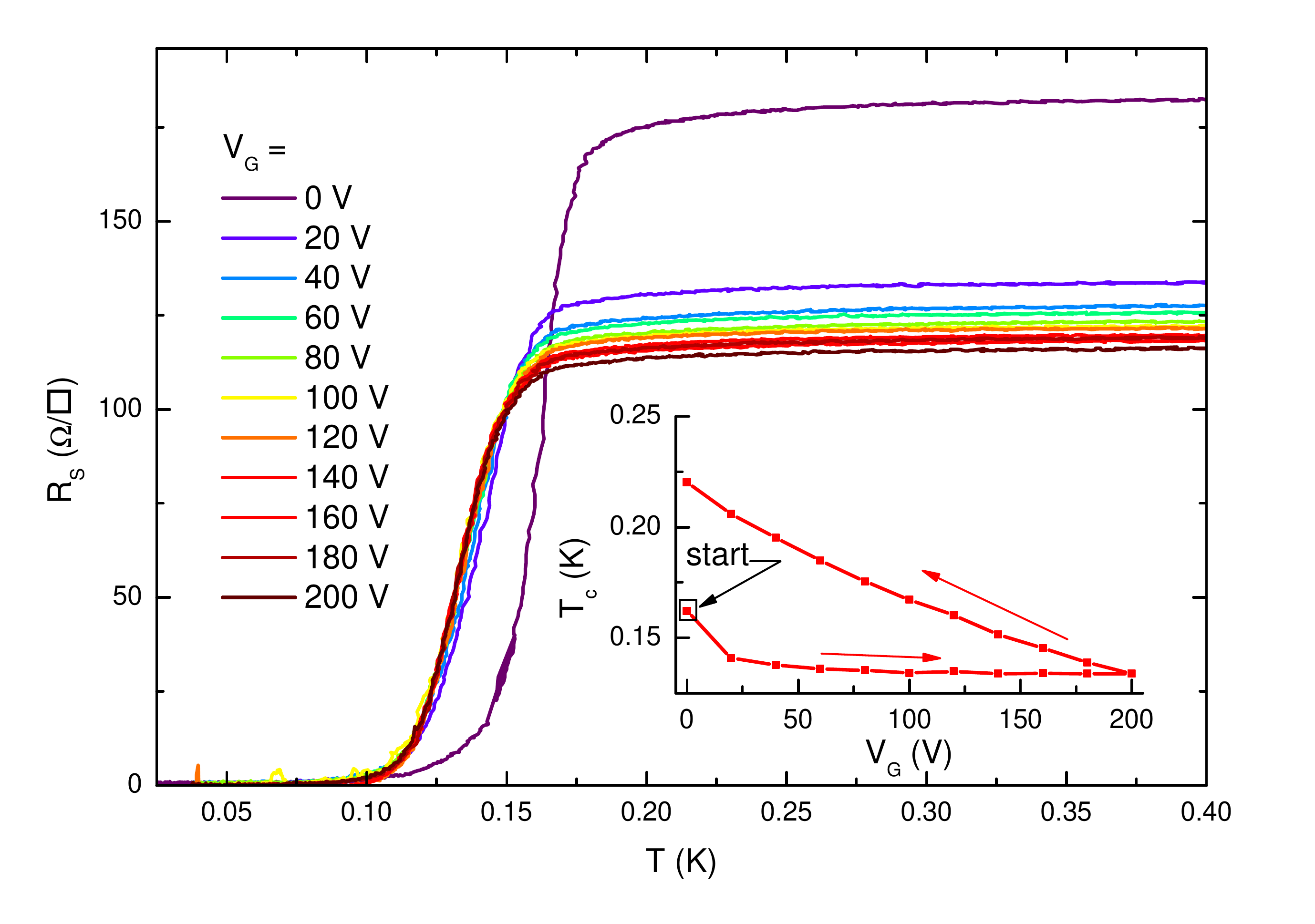}
\caption{Measurement of the resistivity as a function of temperature of the \LAO/\STO sample at different gate voltages during a first positive polarization. Inset : superconducting $T_c$ in the first polarisation to $V_\mathrm{G}^{max}$=+200~V and the following decrease of $V_G$.} 
\label{fs4}
\end{figure}

At lower temperature, the superconducting $T_c$ also saturates during the first positive polarisation (Fig. \ref{fs4}). This is a further indication that indeed all the electronic properties of the 2DEG follow this behavior. This observation is also consistent with the fact that superconductivity is intimately related to High Mobility Carriers (HMC) which set at the edge of the quantum well \cite{biscaras2}. Indeed,  when the un-gated hetero-structure is cooled down, there are spontaneously HMC at the top of the quantum well that induces superconductivity in the 2DEG. During the electrostatic forming, HMC density is roughly constant (see  main text Figure 2b), and so is $T_c$.\\

\begin{figure}[h]
\includegraphics[width=10cm]{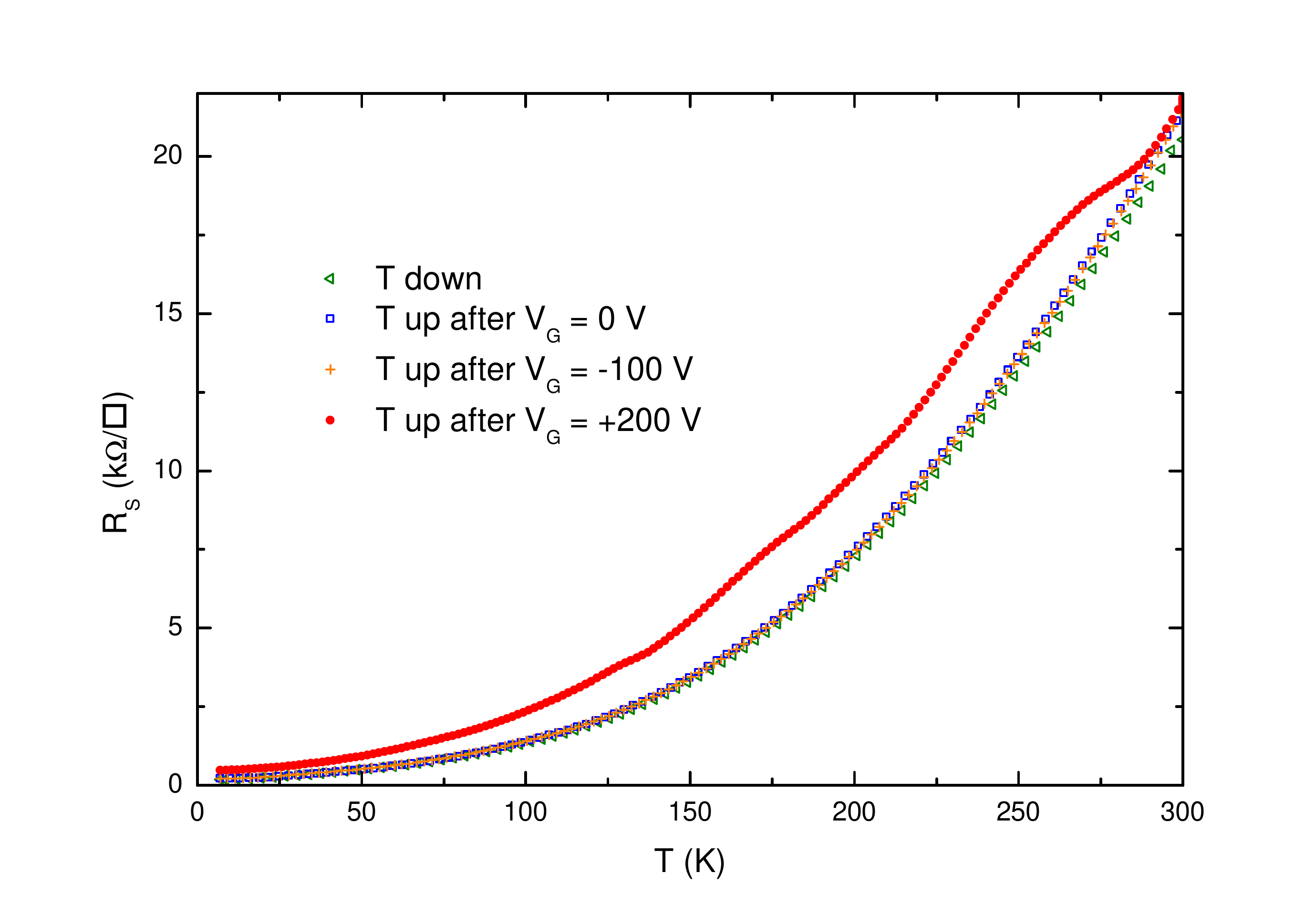}
\caption{Resistance as a function of temperature  of the \LAO/\STO sample measured for different sweep procedures : cool down with $V_\mathrm{G}$=0 (triangle), and warm up with $V_\mathrm{G}$=0 without low temperature polarisation (square), warm up with $V_\mathrm{G}$=0 after a sweep to $V_\mathrm{G}$=-100~V at low temperature (cross) and warm up with $V_\mathrm{G}$=0 after a sweep to $V_\mathrm{G}$=+200~V at low temperature (circle).}
\label{fs5}
\end{figure}
Fig. \ref{fs5} shows the warm-up curve at $V_\mathrm{G}$=0~V after a positive polarisation deep in the irreversible regime. As the temperature is increased, electrons are de-trapped and return to the 2DEG. However, the resistivity steps are less clear than in the \LTO/\STO system and shows numerous release temperatures with various amplitudes : 70~K (almost invisible), 130~K, 170~K, 210~K and 280~K. The 2DEG  goes back to its original state only a high temperature (280~K). The origin of these temperatures and their differences with the \LTO/\STO heterostructure are unknown at the moment.

\end{document}